\documentclass[fleqn,usenatbib]{mnras}

\usepackage{newtxtext,newtxmath}
\usepackage[normalem]{ulem}

\usepackage[T1]{fontenc}
\usepackage{xcolor}

\DeclareRobustCommand{\VAN}[3]{#2}
\let\VANthebibliography\thebibliography
\def\thebibliography{\DeclareRobustCommand{\VAN}[3]{##3}\VANthebibliography}

\usepackage{graphicx}	
\usepackage{amsmath}	

\newcommand{\possis}[1]{\texttt{POSSIS}}

\title{Chemical Distribution of the Dynamical Ejecta in the Neutron Star Merger GW170817}

\author[Shreya Anand et al.]{Shreya Anand,$^{1}$\thanks{E-mail: sanand@caltech.edu}
Peter T. H. Pang,$^{2,3}$\thanks{E-mail: t.h.pang@uu.nl}
Mattia Bulla,$^{4,5,6}$\thanks{E-mail: mattia.bulla@unife.it}
Michael W. Coughlin,$^{7}$
Tim Dietrich,$^{8,9}$
\newauthor
Brian Healy,$^{7}$
Thomas Hussenot-Desenonges,$^{10}$
Theophile Jegou du Laz,$^{11}$
Mansi M. Kasliwal,$^{1}$
\newauthor
Nina Kunert,$^{8}$
Ivan Markin,$^{8}$
Kunal Mooley,$^{1,12}$
Vsevolod Nedora,$^{10}$
Anna Neuweiler$^{8}$
\\
$^{1}$Cahill Center for Astrophysics, California Institute of Technology, Pasadena CA 91125, USA\\
$^{2}$Nikhef, Science Park 105, 1098 XG Amsterdam, The Netherlands\\
$^{3}$Institute for Gravitational and Subatomic Physics (GRASP), Utrecht University, Princetonplein 1, 3584 CC Utrecht, The Netherlands\\
$^{4}$Department of Physics and Earth Science, University of Ferrara, via Saragat 1, I-44122 Ferrara, Italy.\\
$^{5}$INFN, Sezione di Ferrara, via Saragat 1, I-44122 Ferrara, Italy.\\
$^{6}$INAF, Osservatorio Astronomico d’Abruzzo, via Mentore Maggini snc, 64100 Teramo, Italy.\\
$^{7}$School of Physics and Astronomy, University of Minnesota, Minneapolis, Minnesota 55455, USA\\
$^{8}$Institut f\"{u}r Physik und Astronomie, Universit\"{a}t Potsdam, Haus 28, Karl-Liebknecht-Str. 24/25, 14476, Potsdam, Germany\\
$^{9}$Max Planck Institute for Gravitational Physics (Albert Einstein Institute), Am M\"{u}hlenberg 1, Potsdam 14476, Germany\\
$^{10}$IJCLab, Univ Paris-Saclay, CNRS/IN2P3, Orsay, France\\
$^{11}$Division of Physics, Mathematics, and Astronomy, California Institute of Technology, Pasadena, CA 91125, USA\\
$^{12}$Indian Institute of Technology Kanpur, U.P. 208016, India\\
}

\date{Accepted XXX. Received YYY; in original form ZZZ}

\pubyear{2023}

\begin{document}
\label{firstpage}
\pagerange{\pageref{firstpage}--\pageref{lastpage}}
\maketitle

\begin{abstract}
GW170817 and its associated electromagnetic counterpart AT2017gfo continue to be a treasure trove as observations and modeling continue.
Recent precision astrometry of AT2017gfo with the Hubble Space Telescope combined with previous constraints from Very Long Baseline Interferometry (VLBI) constraints narrowed down the inclination angle to 19-25 deg (90\% confidence). 
This paper explores how the inclusion of precise inclination information can reveal new insights about the ejecta properties, in particular, about the composition of the dynamical ejecta of AT2017gfo. Our analysis relies on updated kilonova modeling, which includes state-of-the-art heating rates, thermalization efficiencies, and opacities and is parameterized by $\bar{Y}_{\rm e,dyn}$, the average electron fraction of the dynamical ejecta component. Using this model, we incorporate the latest inclination angle constraint of AT2017gfo into a light curve fitting framework to derive updated parameter estimates. Our results suggest that the viewing angle of the observer is pointed towards the lanthanide-poor ($Y_{\rm e,dyn}\gtrsim0.25$), squeezed polar dynamical ejecta component, which can explain the early blue emission observed in the light curve of AT2017gfo. In contrast to a recent claim of spherical ejecta powering AT2017gfo, our study indicates that the composition of the dynamical ejecta has a strong angular dependence, with a lanthanide-rich ($Y_{\rm e,dyn}\lesssim0.25$), tidal component distributed around the merger plane with a half-opening angle of $35^\circ$. The inclination angle constraint reduces $\bar{Y}_{\rm e,dyn}$ from $0.24$ to $0.22$, with values $0.17\lesssim Y_{\rm e, dyn} \lesssim0.41$ enabling the robust production of $r$-process elements up to the $3^{\rm rd}$ peak in the tidal dynamical ejecta.
\end{abstract}




\section{Introduction}
More than five years since its discovery, the binary neutron star (BNS) merger GW170817 remains the only gravitational-wave event with a definitive electromagnetic (EM) counterpart \citep{abbott2017a}: a low-luminosity short gamma-ray burst \citep{Goldstein2017, Savchenko2017}, a kilonova peaking at ultraviolet to infrared wavelengths \citep{2017gfoAUS, 2017gfoGeminiS, coulter2017, 2017gfoSwift, 2017gfoSpitzer, 2017gfoKilpatrick, 2017gfoMASTER, 2017gfoMcCully, 2017gfoShappee, tanvir2017, 2017gfoJGEM}, and a gamma-ray burst (GRB) afterglow with multi-wavelength emission \citep{Margutti2017, Fong2019, Lamb2019}.
The precise localization of GW170817 to the lenticular galaxy NGC 4993 at 40~Mpc \citep{coulter2017} enabled a detailed study of the energetics of the different outflow components, such as the dynamical ejecta, the disk wind ejecta, and the relativistic jet, responsible for the EM counterparts.

Detectable emission from AT2017gfo lasted for a few weeks post-merger. Due to the initially dominant ultraviolet and blue emission, numerical-relativity simulations indicated the presence of at least two components associated with the kilonova: $\sim0.01 M_{\odot}$~of lanthanide-poor ejecta (i.e.,\ blue component) traveling at an average speed of $\sim$0.3c, and $\sim 0.05 M_{\odot}$~of lanthanide-rich material (i.e.,\ red component) traveling at $\sim$0.1c \citep{arcavi2017, cowperthwaite2017, drout2017, kasen2017, kasliwal2017, nicholl2017, pian2017, smartt2017, soares-santos2017, tanvir2017, valenti2017, villar2017}.
The blue component was interpreted as originating preferentially in the polar regions due to irradiation from neutrinos from a short-lived hypermassive neutron star (HMNS), while the red component to be preferentially equatorial, due to shielding of the neutrinos by the accretion disk \citep{kasen2017,metzger-fernandez2014,metzger2019}. However, ejecta masses and velocities inferred for these two components did not agree well with those predicted by numerical-relativity simulations \citep{Siegel:2019mlp}.
While additional explanations like magnetar-energized wind \citep{metzger2018,yu2018} have been proposed for the kilonova emission, there is no evidence for a long-lived magnetar in GW170817 \citep{pooley2018,margutti2018,makhathini2021,kawaguchi2022,mooley2022}.

The above general conclusions have been drawn from one-dimensional, inclination-independent analyses.
This scenario has changed with the proper motion measurements of the relativistic jet in GW170817, made with the Hubble Space Telescope and very long baseline interferometry \citep{mooley2018-vlbi,ghirlanda2019, mooley2022}, yielding a precise viewing angle constraint of $19<\theta_{\rm obs}<25$ degrees (90\% confidence). Precise measurements of the inclination angle of GW170817 can allow for a more accurate inference of the properties of the associated kilonova, including the ejecta masses, velocities, and composition.

In this work, we present inference results using the tight inclination angle constraint for AT2017gfo and show how the inclusion of inclination angle constraints affects the ejecta parameters and in particular the composition of the dynamical ejecta. We employ a new grid based on an improved version \citep{Bulla2023} of \possis\ ~\citep{Bulla2019}, a three-dimensional Monte Carlo code for modeling the radiation transport in kilonovae. We use idealized outflow properties (geometry and distributions of density and composition) guided by numerical-relativity simulations combined together with the inclination angle constraint.


\begin{figure*}
    \includegraphics[width=0.95\textwidth,clip=True,trim=0pt 130pt 0pt 0pt]{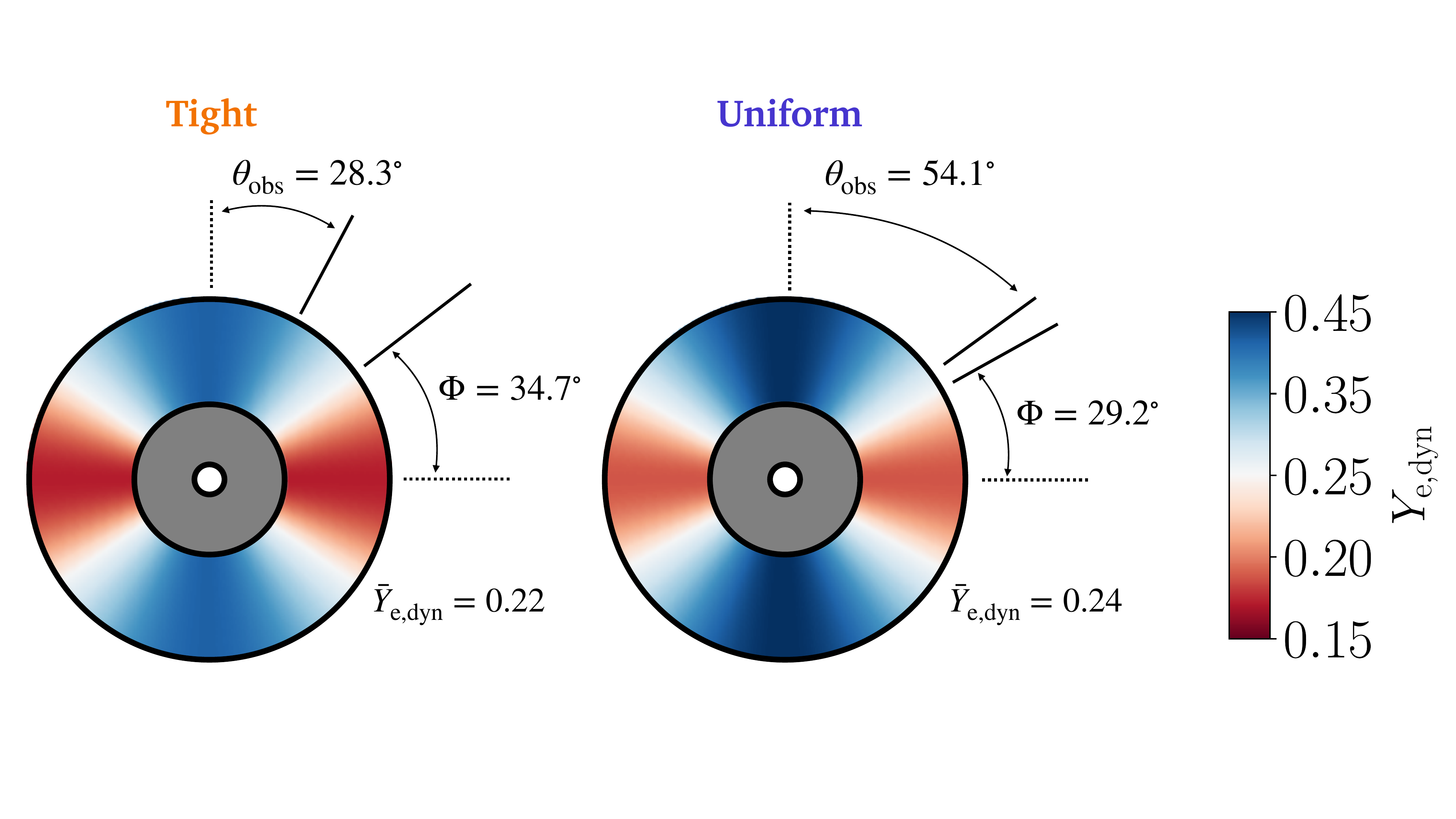}
    \caption{Schematic diagram of the ejecta distribution for AT2017gfo as inferred from the \textsc{Bu2023Ye} grid. The two distinct regions of the dynamical ejecta are represented in blue, corresponding to the lanthanide-poor, squeezed polar dynamical ejecta, and the red, representing the lanthanide-rich, equatorial tidal component. The disk wind ejecta is represented by the grey annulus in the center. The solid lines show the boundary between the lanthanide-rich and lanthanide-poor regions of the dynamical ejecta based on our tight results (left) and our uniform results (right). We quote numbers for $\Phi$, $\bar{Y}_{\rm e, dyn}$, and $\theta_{\rm obs}$. Both runs suggest that the observer is directly viewing the wide-angle, blue polar dynamical ejecta component of AT2017gfo.
    }
    \label{fig:geometry}
\end{figure*}

\section{Methodology} 

\subsection{Kilonova Models} 
\label{sec:KN_modelling}
We make use of the three-dimensional, time-dependent Monte Carlo (MC) radiative transfer code \possis\ ~\citep{Bulla2019,Bulla2023} for simulating kilonova light curves. The code simulates the propagation of MC photon packets as they diffuse out of an expanding medium and interact with matter. Each grid cell in the model is assigned a density $\rho$ (evolved with time assuming homologous expansion), an electron fraction $Y_{\rm e}$ (constant throughout the simulation), and a temperature $T$ (estimated from the mean intensity of the radiation field). At the start of the simulation, MC photon packets are assigned an initial location sampled from the distribution of energy from radioactive-decay of $r$-process nuclei (i.e., depending on the mass and heating rates distribution within the model). The energy available at each time-step is estimated from heating rates \citep{Rosswog2022} and thermalization efficiencies \citep{Barnes2016,Wollaeger2018} that depend on local properties of the ejecta and split equally among MC packets \citep{Lucy1999}. The propagation of MC packets is controlled by the opacity of the ejecta, which is dominated by bound-bound transitions and electron scattering. The most recent version of \possis\ ~\citep{Bulla2023} uses time-dependent opacities that depend on local properties of the ejecta such as $\rho$, $T$, $Y_{\rm e}$, and on the photon frequency/wavelength in the case of bound-bound transitions \citep{Tanaka2020}. We refer the reader to \cite{Bulla2019} and \cite{Bulla2023} for more details about the code.

In this article, we computed a new kilonova grid based on the BNS ejecta model described in \cite{Bulla2023}. The ejecta are modeled following \citet{Kawaguchi2020} and with an idealized geometry assuming two main components: the dynamical ejecta, which is ejected during the merger, and the wind ejecta, which is ejected from the debris disk formed around the central remnant post-merger. The dynamical ejecta, which extend from $0.1$c to $v_{\rm max,dyn}$, where c is the speed of light, have a density profile depending on both the radius $r$ and polar angle $\theta$ as
\begin{equation}
    \rho(r,t,\theta) \propto \begin{cases} \,\sin^2(\theta)\,r^{-4}\,t^{-3} \,\,,\,\,\,\,0.1c\leq r/t \leq 0.4c  \\ \,\sin^2(\theta)\,r^{-8}\,t^{-3} \,\,,\,\,\,\,0.4c\leq r/t \leq v_{\rm max,dyn}  \end{cases}~
    \label{eq:dens}
\end{equation}
and an electron fraction depending on $\theta$ as
\begin{equation}
    Y_{\rm e, dyn}(\theta) = a\,\cos^2(\theta)\,+\,b.
    \label{eq:ye}
\end{equation}
Specifically, we scale the densities and $v_{\rm max,dyn}$ to achieve a desired mass $M_{\rm ej,dyn}$ and mass-weighted averaged velocity $\bar{v}_{\rm ej,dyn}$, respectively, and assume that the parameters $a$ and $b$ are linked as in \cite{Setzer2023} ($a\sim0.71\,b$) and $a$ scaled to obtain a desired mass mean-weighted averaged electron fraction $\bar{Y}_{\rm e, dyn}$. The wind ejecta extend from 0.02 to $v_{\rm max,wind}$, with a density profile $\rho(r)\propto r^{-3}\,t^{-3}$ and a fixed electron fraction $Y_{\rm e,wind}=0.3$. 
The densities and $v_{\rm max,wind}$ are scaled to obtain a desired $M_{\rm ej,wind}$ and mass-weighted averaged velocity $\bar{v}_{\rm ej,wind}$, respectively. If $v_{\rm max,wind}>0.1$c (i.e., the minimum velocity of the dynamical ejecta), we follow \cite{Kawaguchi2020} and assume that the wind ejecta replaces the dynamical ejecta at velocities larger than 0.1c and for angles $\theta<\pi/4$ (see, e.g., their Fig.~2, bottom panel).\\

The new kilonova grid, hereafter referred to as \textsc{Bu2023Ye}, depends on five free parameters in addition to the viewing angle $\Theta_{\rm obs}$: 
$M_{\rm ej,dyn}$, $\bar{v}_{\rm ej,dyn}$, $\bar{Y}_{\rm e,dyn}$, $M_{\rm ej,wind}$, and $\bar{v}_{\rm ej,wind}$. 
Values for these parameters are chosen to cover the expected ranges from numerical-relativity simulations of binary neutron star mergers \citep[e.g.][]{Radice2018,Nedora2021}: $M_{\rm ej,dyn}=[0.001,0.005,0.010, 0.020]\,M_\odot$, $\bar{v}_{\rm ej,dyn}=[0.12,0.15,0.20,0.25]$c, $\bar{Y}_{\rm e,dyn}=[0.15,0.20,0.25,0.3]$, $M_{\rm ej,wind}=[0.01,0.05,0.09,0.13]\,M_\odot$ and $\bar{v}_{\rm ej,wind}=[0.03,0.05,0.10,0.15]$c. This results in a total of $4\times4\times4\times4\times4=1024$ configurations, corresponding to $11264$ different kilonova light curves when counting the $N_{\rm obs}=11$ different viewing angles $\theta_{\rm obs}$ (equally spaced in cosine from a face-on / jet-axis to an edge-on / merger-plane view of the system). Each simulation employs a number $N_{\rm ph}=10^6$ of MC photon packets, $N_{\rm times}=100$ time-steps from $0.1$ to $30$~d after the merger (logarithmic binning of $\Delta \log t=0.025$) and $N_{\rm \lambda}=1000$ wavelength bins from $500$~\AA{} to $10\,\mu$m (logarithmic binning of $\Delta \log\lambda=0.0023$). 

For comparison, we also make use of the grid introduced in \citet{Dietrich2020}, parameterized by $M_{\rm ej, dyn}$, $M_{\rm ej, wind}$, $\Phi$, and $\theta_{\rm obs}$. This kilonova grid uses an earlier version of \possis\ ~\citep{Bulla2019} with analytic functions for the opacities and simpler prescriptions for heating rates and thermalization efficiencies. In the rest of the paper, the model built upon the old grid, \cite{Dietrich2020}, is referred to as \textsc{Bu2019lm}.

Fig.~\ref{fig:geometry} shows a sketch of the ejecta for the best-fit model found in this work. We note that the constraint on the averaged $\bar{Y}_{\rm e,dyn}$, coupled with the $Y_{\rm e,dyn}(\theta)$ angular dependence in Eq.~\eqref{eq:ye}, leads to a constraint on the chemical distribution of the dynamical ejecta and specifically on the half-opening angle $\Phi$ of the lanthanide-rich ($Y_{\rm e}<0.25$) region, see below. 

\subsection{Bayesian analysis using \texttt{NMMA}}
\label{sec:nmma}
To analyse AT2017gfo, we use the Nuclear physics and Multi-Messenger Astronomy framework \texttt{NMMA}~\citep{Dietrich2020, Pang2022}\footnote{\url{https://github.com/nuclear-multimessenger-astronomy/nmma}} that allows us to perform joint Bayesian inference of multi-messenger events containing gravitational waves, kilonovae, supernovae, and GRB afterglows.

To perform Bayesian analysis on the light curve data using the grid of predicted light curves, an interpolation scheme is required. Here, we employ a feed-forward neural network (NN) to predict the kilonova light curves based on the input parameters~\citep{Almualla:2021znj}. We used a NN to create a continuous mapping between merger parameters and light curve eigenvalues. The NN architecture begins with an input layer followed by one dense layer with 2048 neurons. A dropout layer subsequently removes 60\% of the dense layer's outputs before connecting to the output layer, yielding ten eigenvalues. This regularized, wide and shallow NN approximates a Gaussian process (see e.g., \citealt{2017arXiv171100165L}) in a fraction of its runtime. We used 90\% of learning set examples to train the models over 100 epochs, reserving the remaining 10\% for validation.

The inference of kilonovae is based on the AB magnitude for a specific filter $j$, $m^{j}_{i}(t_i)$. The measurements are given as a time series at times $t_i$ with a corresponding statistical error $\sigma^{j}_{i} \equiv \sigma^{j}(t_i)$. 
The likelihood function $\mathcal{L}(\vec{\theta})$ is given by~\cite{Pang2022}:
\begin{equation}
\begin{aligned}
    \mathcal{L}(\vec{\theta}) \propto \exp\left(- \frac{1}{2}\sum_{ij}\frac{\left(m^{j}_{i} - m^{j, {\rm est}}_{i}(\vec{\theta})\right)^2}{(\sigma^j_i)^2 + (\sigma_{\rm sys})^2}\right)
    \propto \exp\left(-\frac{1}{2}\chi^2(\vec{\theta})\right),
\end{aligned}
\end{equation}
where $m^{j, \rm{est}}_{i}(\vec{\theta})$ is the estimated AB magnitude for the parameters $\vec{\theta}$ using the interpolation scheme. $\sigma_{\rm sys}$ is the additional error budget included to account for the systematic uncertainty within the kilonova modeling. Such a likelihood is equivalent to including an additional shift to the light curve by $\Delta m$, and marginalizing it with a normal distribution with mean of $0$ and variance of $\sigma^2_{\rm sys}$. Similar to \cite{Pang2022} and motivated by the study of \cite{Heinzel2021}, $\sigma_{\rm sys}$ is taken to be $1{\rm mag}$.

The nested sampling algorithm implemented in \texttt{PyMultiNest} \citep{Buchner:2014nha, Feroz2009} is used for exploring the likelihood landscape and obtaining posterior samples.

\section{Parameter Inference}
\label{sec:pe}

\begin{figure}
    \centering
    \includegraphics[width=0.5\textwidth]{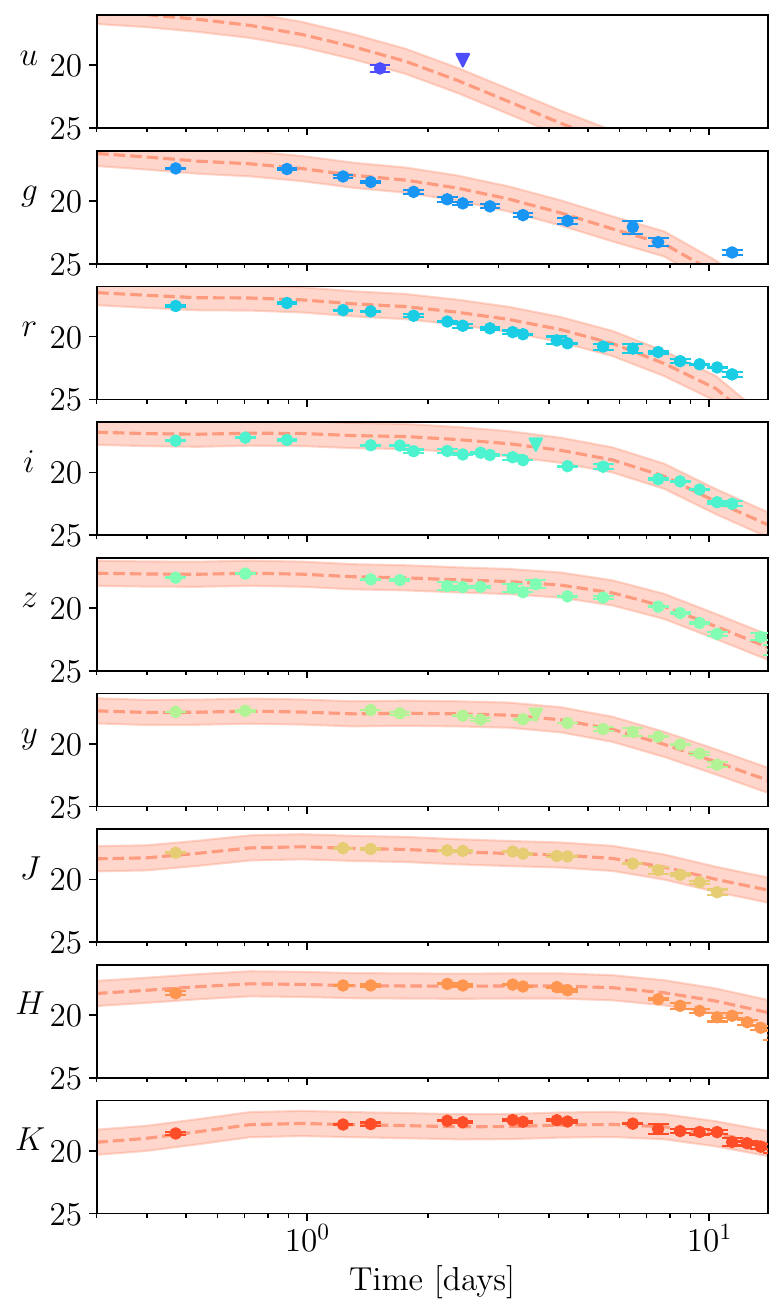}
    \caption{Best-fit light curve model from \textsc{Bu2023Ye} tight run on top of the AT2017gfo data. The data points are mostly within the 1mag systematics uncertainty band from the best-fit light curves, with $\chi^2 / {\rm d.o.f.} = 0.447$.}
    \label{fig:lightcurve}
\end{figure}

\begin{figure*}
    \hspace{-0.5cm}
    \includegraphics[width=0.55\textwidth]{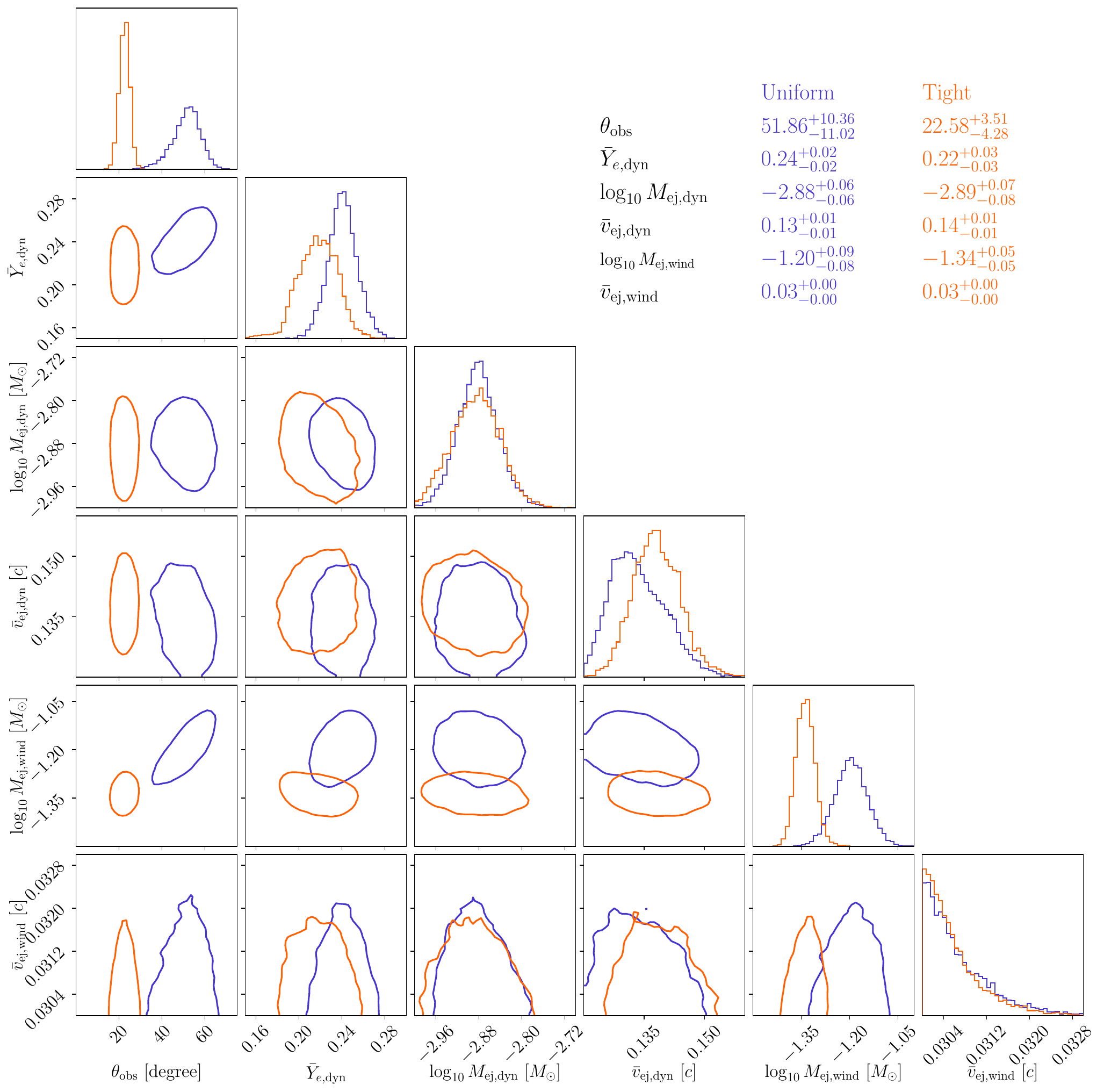}\includegraphics[width=0.45\textwidth]{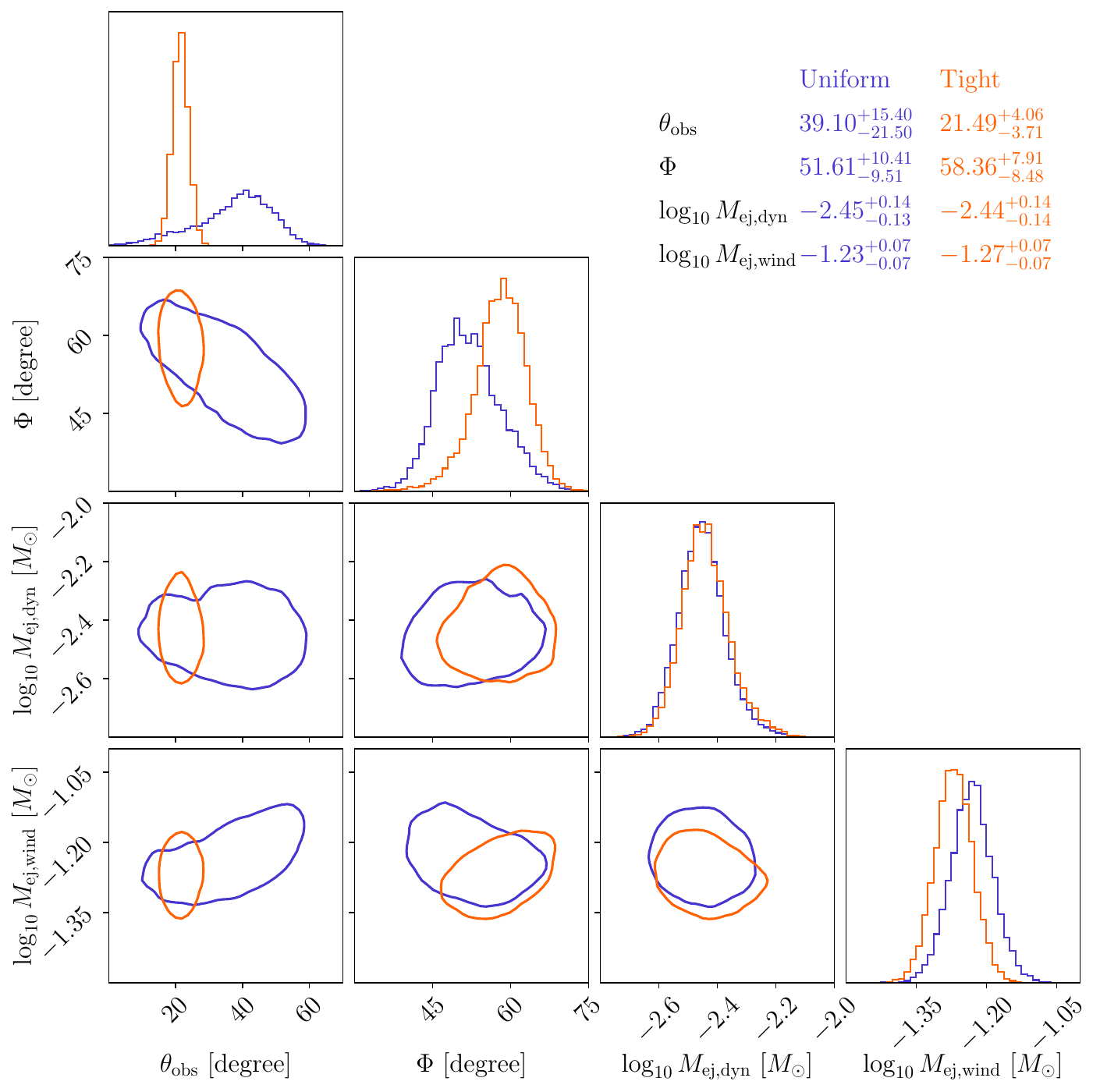}
    \caption{Comparison between the inference results for the \textsc{Bu2023Ye} (\textit{left}) and the \textsc{Bu2019lm} (\textit{right}) models with the inclusion of a tight Gaussian prior (orange) and a uniform inclination prior (blue). The median values (with $95\%$ credibility uncertainty) of parameters are also shown.}
    \label{fig:corner_comparison}
\end{figure*}

As described in Sec.~\ref{sec:nmma}, Bayesian analyses are performed using \textsc{Bu2019lm} and \textsc{Bu2023Ye}. For both models, we have considered two sets of priors. The \textit{uniform} prior has a uniform prior on the viewing angle $\theta_{\rm obs}$. Instead, the \textit{tight} prior, applying the constraint in \cite{mooley2018-vlbi,mooley2022}, has a Gaussian prior on the viewing angle with a mean of $21.3^\circ$ and a standard deviation of $2.5^\circ$. In contrast to the uniform prior, our tight prior is constrained by the precise inclination measurement of GW170817 and its associated uncertainty, which can be used to gauge how precise inclination constraints can influence the inference of other intrinsic ejecta properties. The priors on the rest of the parameters are shown in Tab.~\ref{tab:prior}.
The best-fit light curves of the \textsc{Bu2023Ye} \textit{tight} run are shown in Fig.~\ref{fig:lightcurve}. 
The posteriors and the best-fit values with their corresponding $\chi^2$ are shown in Fig.~\ref{fig:corner_comparison} and Tab.~\ref{tab:bestfit_params}, respectively. 

\begin{table}
    \centering
    \begin{tabular}{l|l|l}
    \hline\hline
        Parameter & \textsc{Bu2023Ye} & \textsc{Bu2019lm}\\
        \hline
        $\log_{10}(M_{\rm ej, dyn} \ [M_{\odot}])$ & $\mathcal{U}(-3, -1.7)$ & $\mathcal{U}(-3, -1)$\\
        $\bar{v}_{\mathrm{ej, dyn}} \ [c]$ & $\mathcal{U}(0.12, 0.25)$ & $-$\\
        $\bar{Y}_{\mathrm{e, dyn}}$ & $\mathcal{U}(0.15, 0.30)$ & $-$\\
        $\log_{10}(M_{\rm ej, wind} \ [M_{\odot}])$ & $\mathcal{U}(-2, -0.89)$ & $\mathcal{U}(-3, -0.5)$\\
        $\bar{v}_{\mathrm{ej, wind}} \ [c]$ & $\mathcal{U}(0.03, 0.15)$ & $-$\\
        $\Phi \ [^\circ]$ & $-$ & $\mathcal{U}(15, 75)$\\
        \hline\hline
    \end{tabular}
    \caption{Parameter priors used for the Bayesian analyses. $\mathcal{U}(a, b)$ refers to an uniform prior in the interval $[a, b)$. This table excludes the priors on the inclination angle, which we describe in Sec.~\ref{sec:pe}.}
    \label{tab:prior}
\end{table} 

For the analyses using the \textsc{Bu2019lm} model, the inclusion of the tight prior on the viewing angle noticeably changes the posterior and the best-fit value of the viewing angle and the opening angle of the tidal dynamical ejecta, $\Phi$. On the other hand, the rest of the parameters are only marginally affected by such an inclusion. Such a phenomenon can be explained by the simpler geometry in the \textsc{Bu2019lm} model and the opening angle's major role in the anisotropic geometry of the ejecta. The inclusion of the viewing angle constraint is causing an increase in the opening angle ($49.74^\circ \to 58.26^\circ$) and decrease in the disk wind ejecta mass ($0.060 M_\odot \to 0.056 M_\odot$).

Due to the more complicated ejecta geometry in the \textsc{Bu2023Ye} model, the posteriors of multiple parameters are significantly altered by the changes in the viewing angle prior. The \textsc{Bu2023Ye} tight run converges at a best-fit value $\bar{Y}_{\mathrm{e, dyn}}=0.22$, which is lower than the result for the \textsc{Bu2023Ye} uniform run ($\bar{Y}_{\mathrm{e, dyn}}=0.24$). Aside from the inferred inclination angle, which changes from 54$^\circ$ (\textsc{Bu2023Ye} uniform) to $28^\circ$ (\textsc{Bu2023Ye} tight), the disk wind ejecta mass also changes between the two runs, from $\sim 0.06 M_\odot$ to $\sim 0.05 M_\odot$ with reduced uncertainty. This change can be accounted for by the correlation between the viewing angle and the disk wind ejecta mass (see 2D contour plot in Fig.~\ref{fig:corner_comparison}).

\begin{figure}
    \centering
    \includegraphics[width=0.5\textwidth]{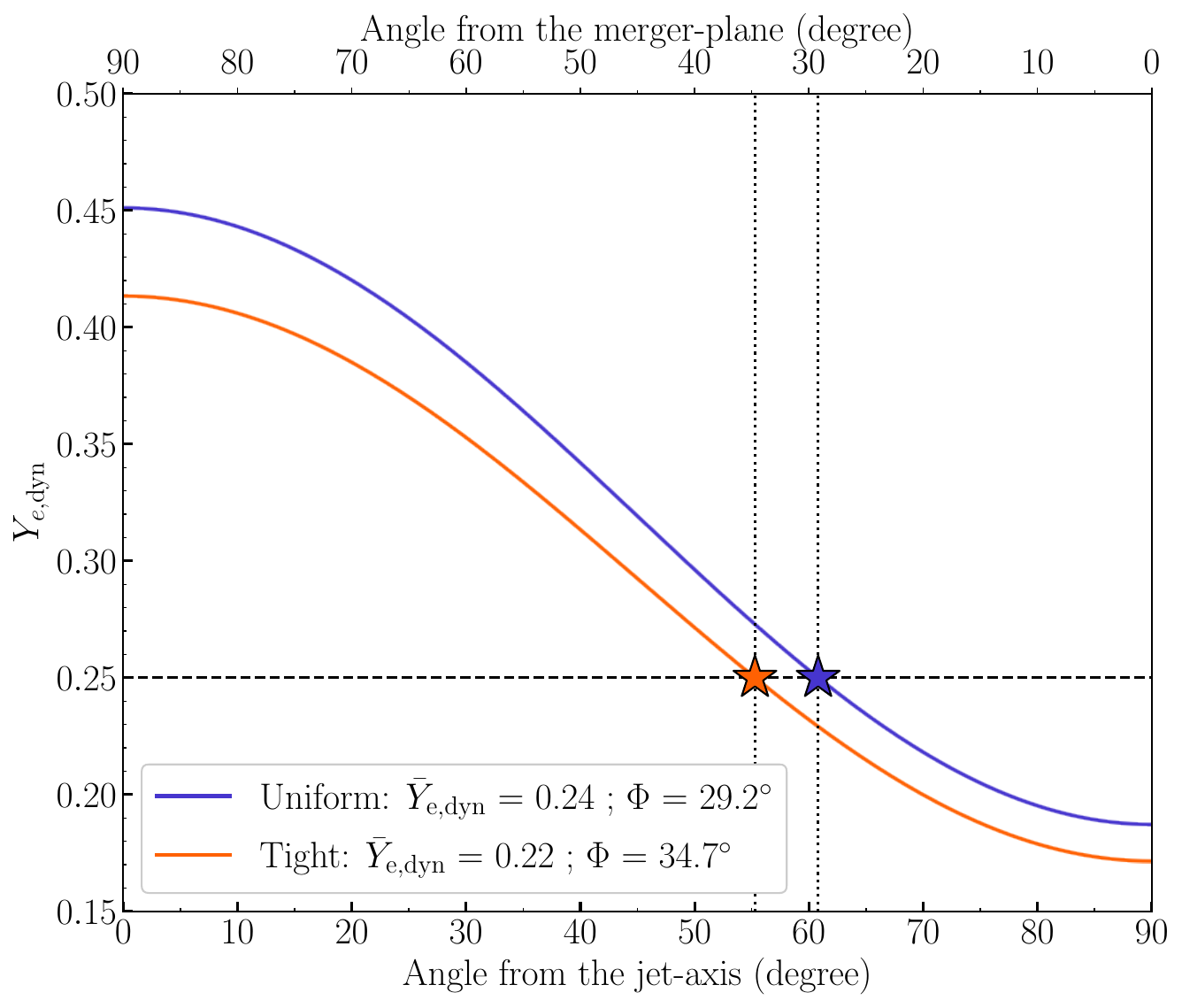}
    \caption{Constraints on the extension of the lanthanide-poor squeezed-polar ($Y_{\rm e,dyn}\ge0.25$) and lanthanide-rich tidal-equatorial ($Y_{\rm e,dyn}<0.25$) dynamical ejecta for the \textsc{Bu2023Ye} model. The viewing angle dependence used in \possis\ ~(Eq.~\ref{eq:ye}), together with the averaged values $\bar{Y}_{\rm e,dyn}=0.22\,(0.24)$ inferred from the uniform (tight) run suggests that the half-opening angle of the lanthanide-rich component is $\Phi=34.7^\circ$ ($29.2^\circ$).}
    \label{fig:Yedyn}
\end{figure}

Our latest results from the \textsc{Bu2023Ye} model suggest that a smaller quantity of ejecta is contributing to the total kilonova than previously inferred by \textsc{Bu2019lm} \citep{Dietrich2020, Pang2022}. In both cases, with and without the inclusion of the inclination angle constraint, our new inferred value for the dynamical ejecta mass is $M_{\mathrm{ej, dyn}}\approx0.001 M_{\odot}$, while from the \textsc{Bu2019lm} models it was estimated to be $M_{\mathrm{ej, dyn}}\approx0.002 M_{\odot}$. Moreover, a smaller quantity of wind ejecta mass is favored from the tight run with our new models. Amongst all of the different runs, the \textsc{Bu2023Ye} tight run differs significantly in its inference of $\bar{Y}_{\rm e,dyn}$. Figure~\ref{fig:Yedyn} shows the relationship between $\theta_{\rm obs}$, $Y_{\rm e,dyn}$, and $\Phi$ and highlights how the electron fraction in the dynamical ejecta is constrained to vary in the range $0.17\lesssim Y_{\rm e, dyn} \lesssim0.41$ in the tight run. The intersection of each $Y_{\mathrm{e, dyn}} (\theta_{\rm obs})$ with $Y_{\mathrm{e, dyn}}=0.25$, or the boundary between the lanthanide-rich and lanthanide-poor regions of the dynamical ejecta component dictates the corresponding value of $\Phi$. Thus, we calculate a value of $34.7^\circ$ for the \textsc{Bu2023Ye} tight run. Together with the inferred total mass of $1.35\times10^{-3} M_{\odot}$ and density profile (Eq.~\ref{eq:dens}) of the dynamical ejecta, this $\Phi$ value implies that the squeezed-polar component has a mass of $0.24\times10^{-3} M_{\odot}$ while the tidal-equatorial component a mass of $1.11\times10^{-3} M_{\odot}$, nearly five times larger.

\begin{table*}
    \centering
    \begin{tabular}{ccccccccccc}
    \hline\hline
     model & prior & $\log_{10}(M_{\mathrm{ej, dyn}})$ & $\bar{v}_{\mathrm{ej, dyn}}$ & $\bar{Y}_{\rm e,dyn}$ & $\log_{10}(M_{\mathrm{ej, wind}})$ & $\bar{v}_{\mathrm{ej, wind}}$ & $\Phi$ & $\theta_{\rm obs}$ & $\chi^2 / {\rm d.o.f.}$\\
     units & - & $\log_{10}(M_{\odot})$ & $(c)$ & - & $\log_{10}(M_{\odot})$ & $(c)$ & ($^{\circ}$) & ($^{\circ}$) & - \\
     \hline
      \textsc{Bu2023Ye} & uniform & $-2.89$ & $0.13$ & $0.24$ & $-1.19$ & $0.03$ & -- & $54.14$ & $0.380$\\
      \textsc{Bu2023Ye} & tight & $-2.87$ & $0.14$ & $0.22$ & $-1.32$ & $0.03$ & -- & $28.28$ & $0.447$\\
      \textsc{Bu2019lm} & uniform & $-2.43$ & - & - & $-1.22$ & - & $49.74$ & $43.72$ & $0.461$\\
      \textsc{Bu2019lm} & tight & $-2.48$ & - & - & $-1.25$ & - & $58.26$ & $25.96$ & $0.480$\\
      \hline\hline
    \end{tabular}
    \caption{The best-fit parameters for each inference setting with their corresponding $\chi^2$ per degrees of freedom.}
    \label{tab:bestfit_params}
\end{table*}

\begin{figure}
    \centering
    \includegraphics[width=0.49\textwidth]{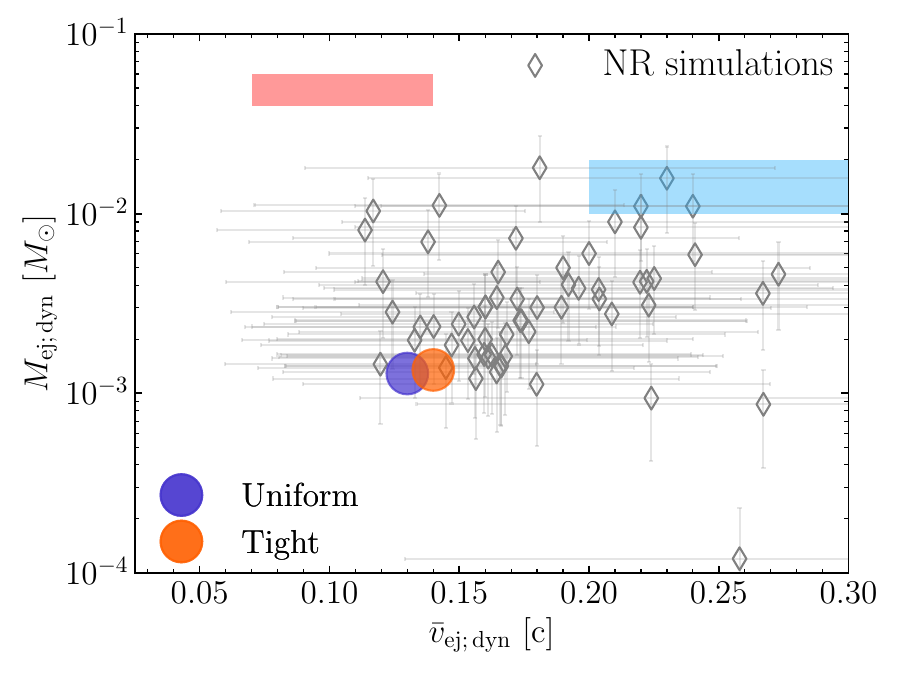}
    \includegraphics[width=0.48\textwidth]{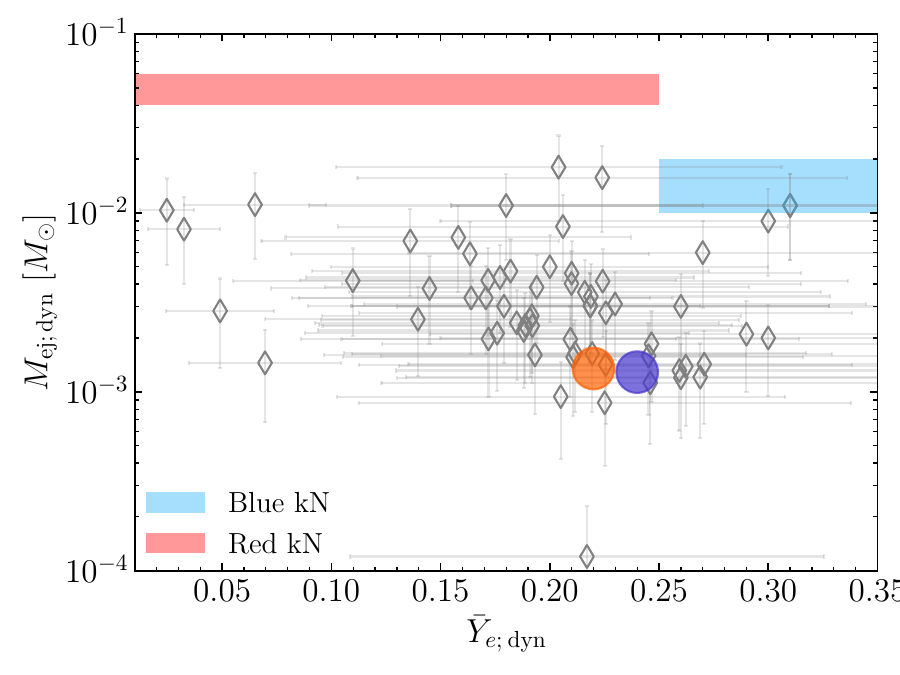}
    \caption{
    Predictions for the dynamical ejecta properties: mass, velocity, and electron fraction (as listed in Tab.~\ref{tab:bestfit_params}) alongside the data from numerical-relativity simulations of BNS mergers (see text for details) and previous expectations for AT2017gfo from \protect\cite{Siegel:2019mlp} (shown as blue and red boxes for the two components of ejecta considered). 
    Our measurements for the dynamical and wind ejecta masses and velocities (colored circles) are compared to numerical relativity simulations from \protect\cite{Nedora2022}.
    }
    \label{fig:Nedora_NR}
\end{figure}

\section{Discussion}

Our results provide us with new insights into the angular distribution of ejecta composition of AT2017gfo. Our analysis of the light curve of AT2017gfo with the updated \textsc{Bu2023Ye} model with time-dependent opacities is more robust than previous studies relying on the \textsc{Bu2019lm} model. Here we first discuss how the inclusion of the inclination angle constraint modifies the geometry and the composition of the ejecta relative to a uniform inclination angle case.

Compared to \textsc{Bu2023Ye} uniform, the \textsc{Bu2023Ye} tight run also favors a slightly more extended (i.e.,\ wider-angle) equatorial dynamical ejecta component. However, we find a smaller value of $\Phi$ ($34.7^\circ$) compared to \textsc{Bu2019lm}. Since $\Phi$ controls the opening angle of the tidal dynamical ejecta, our results point towards a wide-angle squeezed-polar component and a narrow-angle tidal-equatorial component to the dynamical ejecta. The inclination angle constraint further solidifies that the observer's line-of-sight is pointed toward the squeezed polar dynamical ejecta, which contributes to the observed early blue light curve and affects the kilonova color evolution. As shown in Fig.~\ref{fig:contributions}, a good fraction ($\gtrsim50$\%) of the radiation escaping at early times ($\lesssim2$\,d) to an observer at $\theta_{\rm obs}\sim28^\circ$ originates in the lanthanide-poor squeezed-polar component, which is able to explain the early blue emission in AT2017gfo (see Fig.~\ref{fig:lightcurve}) without the need of additional sources as suggested in the literature \citep[e.g.][]{arcavi2017,kasliwal2017,arcavi2018,metzger2018,piro2018}. Alternatively, for a kilonova with identical geometry and composition to what we infer from this study to appear \emph{without} an early blue component to its light curve, the viewing angle would have to be $\gtrsim55$ deg. An AT2017gfo-like kilonova would likely exhibit a blue-to-red color evolution for most observer viewing angles.

In addition to the analysis of observational data to reveal the properties of the ejecta, ab-inito numerical-relativity simulations with advanced input physics can also be used to predict the properties of the ejecta and to link these properties to the BNS source parameters; however, for an accurate estimation of the ejecta electron fraction, sophisticated neutrino transport schemes are required. In Fig.~\ref{fig:Nedora_NR}, we show the mass, average velocity, and electron fraction of dynamical ejecta from a large sample of numerical-relativity simulations that included neutrino transport. The sample is based on the data collected in \cite{Nedora2022} and corresponds to the \texttt{M0RefSet} combined with \texttt{M0/M1Set} there. In these simulations, both neutrino emission and absorption were included, albeit in a different manner. In simulations with M0 neutrino scheme, neutrinos were assumed to move radially, depositing only energy to the fluid, while in M1 scheme, neutrino motion was unconstrained, and both energy and momentum deposition was included. These simulations were collected from the following works: \cite{Radice2018,Sekiguchi:2015dma,Sekiguchi:2016bjd,Vincent:2019kor,Perego:2019adq,Nedora:2019jhl,Bernuzzi:2020txg,Nedora2021}\footnote{See also more recent works \citep[e.g.,][]{Fujibayashi:2022ftg}}. 
Square red and blue boxes in Fig.~\ref{fig:Nedora_NR} correspond to the expectations derived from the analysis of multiwavelength data of AT2017gfo \citep[e.g.,][]{villar2017,Perego:2017wtu} compiled by \citet{Siegel:2019mlp}. Notably, these kilonova models were largely semi-analytical or one-dimensional. With a more sophisticated, anisotropic radiation transport model that takes into account ejecta composition and structure, we obtain properties of the dynamical ejecta that are more consistent with what numerical-relativity simulations suggest (see cyan and orange circles in the figure). Specifically, the inferred ejecta velocities and electron fractions are now closer to the average values obtained with numerical-relativity simulations. This highlights the importance of multi-dimensional radiative transfer simulations for kilonova modelling \citep[see also][]{Kawaguchi2018,Korobkin2021,Collins2023,Kedia2023,Shingles2023}.

In addition to the agreement with respect to the ejecta mass, the ejecta velocity, and also the electron fraction, we also find large consistency between the half-opening angle of about $35^\circ$ and numerical-relativity predictions, e.g., employing the prediction of \cite{Nedora2022} for $\Phi$ using the mean value of $q$ and $\tilde{\Lambda}$ of~\cite{Dietrich2020}, we find $\Phi \sim 32.7^\circ$, which is consistent with our prediction (Fig.~\ref{fig:Yedyn}). 

The low inferred average electron fraction of $\bar{Y}_{\rm e,dyn}=0.22$ hints at the formation of heavier species in the KN ejecta. Our inferred $\bar{Y}_{\rm e,dyn}=0.22$ for the tight prior run reveals a dynamical ejecta composition that varies between $Y_{\rm e,dyn}=0.17$ and $Y_{\rm e,dyn}=0.41$. Based on the nuclear reaction network calculations from \citet{Lippuner2015} and \citet{Rosswog2018}, an ejecta composition of $Y_{\rm e,dyn}\lesssim0.20$ can well reproduce the solar $r$-process abundance pattern at $2^{\rm nd}$ peak and beyond. Thus, by inferring that the distribution of electron fractions extends down to $Y_{\rm e,dyn}=0.17$, our work insinuates that AT2017gfo likely produced $3^{\rm rd}$ peak elements consistent with the solar $r$-process abundance pattern.

Finally, we can also compare our results with a recent study of \cite{Sneppen2023}, which suggested that the broad absorption feature observed in the optical spectra of AT2017gfo and attributed to strontium \citep{Watson2019} is consistent with a kilonova that is highly spherical at early times. Namely, the pole-to-equator variation in the density and $Y_{\rm e,dyn}$ distribution is constrained to be small and shifted towards relatively high values needed to produce strontium. Our results indicate the opposite, suggesting that a strong pole-to-equator variation in $Y_{\rm e}$ and lanthanide-rich compositions close to the equator are required to reproduce the observed light curve evolution.

\section{Conclusions}

In this work, we use a tight inclination constraint for GW170817/AT2017gfo from recent VLBI observations \citep{mooley2022} and explore how this affects parameter inference on the ejecta, with a particular focus on the composition/chemical distribution within the dynamical ejecta component. We present a new grid of 1024 kilonova models computed with the three-dimensional Monte Carlo radiative transfer code \possis/ that depends on five parameters: the mass ($M_{\mathrm{ej, dyn}}$), averaged velocity ($\bar{v}_{\mathrm{ej, dyn}}$) and averaged electron fraction ($\bar{Y}_\mathrm{{e, dyn}}$) of the dynamical ejecta, and the mass ($M_{\mathrm{ej, wind}}$) and averaged velocity ($\bar{v}_{\mathrm{ej, wind}}$) of the post-merger disk-wind ejecta. Guided by numerical-relativity simulations, an angular dependence of both density and electron fraction is assumed for the dynamical ejecta. Employing this grid, we train a feed-forward NN that allows us to compute generic light curves to perform Bayesian parameter estimation. The main results can be summarized as follows:
\begin{itemize}
    \item The best-fit model to AT2017gfo when adopting a tight prior on the inclination angle corresponds to $M_{\mathrm{ej, dyn}}=1.3\times10^{-3}\,M_\odot$, $\bar{v}_{\mathrm{ej, dyn}}=0.14$c, $\bar{Y}_\mathrm{{e, dyn}}=0.22$, $M_{\mathrm{ej, wind}}=4.8\times10^{-2}\,M_\odot$ and $\bar{v}_{\mathrm{ej, wind}}=0.03$c. Compared to a run with no constraints on the inclination angle, the inferred masses are higher and the averaged electron fraction lower.
    \item The inferred averaged electron fraction of $\bar{Y}_\mathrm{{e, dyn}}=0.22$ points to a strong angular variation of composition within the dynamical ejecta, with the electron fraction varying from $Y_{\rm e, dyn}\sim0.17$ in the merger plane to $Y_{\rm e, dyn}\sim0.41$ at the pole. The lanthanide-rich ($Y_{\rm e}\lesssim0.25$), tidal component is distributed around the merger plane with a half-opening angle $\Phi=35^\circ$, while the rest of the dynamical ejecta is lanthanide-poor ($Y_{\rm e}\gtrsim0.25$).
    \item The viewing angle of GW170817/AT2017gfo is pointed towards the lanthanide-poor (blue), squeezed polar dynamical ejecta, which can explain the early light curves of the observed kilonova without the need for additional powering sources.
    \item The properties inferred for the dynamical ejecta are in good agreement with those predicted by numerical-relativity simulations, alleviating the tension found previously when adopting one-dimensional and/or semi-analytical kilonova models \citep[see e.g.][for a review]{Siegel:2019mlp}.
    \item The large extent of a lanthanide-rich component within the dynamical ejecta is in contrast with a recent claim, based on the analysis of the strontium P Cygni spectral feature in AT2017gfo, of a spherical kilonova and a small pole-to-equator variation of $Y_{\rm e}$ \citep{Sneppen2023}.
    \item The range of electron fraction values inferred for the dynamical ejecta, going as low as $Y_{\rm e, dyn}\sim0.17$, leads to the production of $r-$process nuclei up to the $3^{\rm rd}$ $r-$process peak \citep[e.g.][]{Lippuner2015,Rosswog2018}. 
\end{itemize}

Our study highlights the importance of using tight inclination constraints and state-of-the-art multi-dimensional kilonova modeling for parameter inference of the ejecta properties. Furthermore, we underscore our ability to probe the nucleosynthesis of heavy elements using just the early-time kilonova photometry alone. Given the sensitivity of current ground-based gravitational-wave detectors to distant neutron star mergers, detailed spectroscopy of associated kilonova counterparts to gravitational-wave sources may not always be tenable. With a precise inclination angle measurement from an associated on-axis GRB or via other means, one can employ this methodology to infer the chemical distribution of any kilonova's ejecta using well-sampled early light curve data. This approach, paired with spectroscopic studies of nearby KNe, can enable a better understanding of the nucleosynthetic yields from a population of kilonovae.

\section*{Acknowledgements}
SA acknowledges support from the National Science Foundation GROWTH PIRE grant No. 1545949.
PTHP is supported by the research program of the Netherlands Organization for Scientific Research
(NWO).
MWC and BH acknowledge support from the National Science Foundation with grant numbers PHY-2010970 and OAC-2117997.
TD acknowledge funding from the EU Horizon under ERC Starting Grant, no.\ SMArt-101076369. TD and AN acknowledge support from the Deutsche
Forschungsgemeinschaft, DFG, project number DI 2553/7. 
TD and VN acknowledge support through the Max Planck Society funding the Max Planck Fellow group 
`Multi-messenger Astrophysics of Compact Binaries'. The simulations were performed on the national supercomputer HPE Apollo Hawk at the High-Performance Computing (HPC) Center Stuttgart (HLRS) under the grant number GWanalysis/44189 and on the GCS Supercomputer SuperMUC\_NG at the Leibniz Supercomputing Centre (LRZ) [project pn29ba]. 

\section*{Data Availability}

The datasets used and the software used for the analysis can be found at (\url{https://github.com/nuclear-multimessenger-astronomy/nmma}). The interpolation models can be found on Zenodo (\url{https://doi.org/10.5281/zenodo.8039909}).
The raw data will be made public once the paper is accepted to the journal.
 



\bibliographystyle{mnras}
\bibliography{references} 




\appendix

\section{Flux contribution from different ejecta components}

\begin{figure}
    \centering
    \includegraphics[width=\columnwidth]{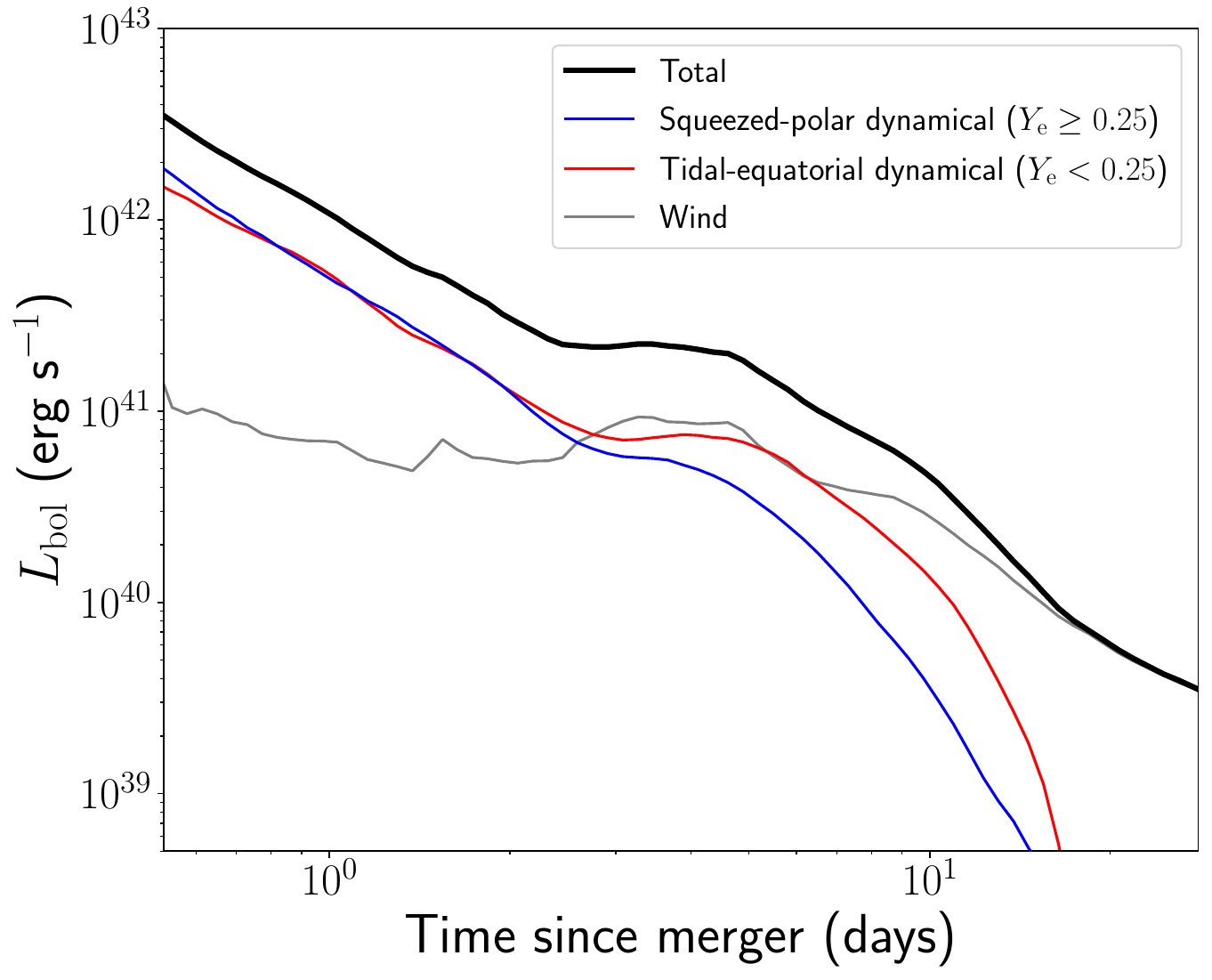}
    \caption{Contribution to the total bolometric light curve (black) of the squeezed-polar dynamical ($Y_{\rm e}\ge 0.25$, blue), tidal-equatorial dynamical ($Y_{\rm e}<0.25$, red) and wind (grey) ejecta components. light curves have been computed with a dedicated \possis\ ~simulation with $N_{\rm ph}=10^7$ using the best-fit parameters of the \textsc{Bu2023Ye} tight run and for an observer viewing angle of $\theta_{\rm obs}=28^\circ$.}
    \label{fig:contributions}
\end{figure}

Fig.~\ref{fig:contributions} shows the contribution to the bolometric light curves of the squeezed-polar dynamical ($Y_{\rm e}\ge 0.25$, blue), the tidal dynamical ($Y_{\rm e}<0.25$, red) and the wind (grey) ejecta components. light curves refer to a dedicated ($N_{\rm ph}=10^7$) \textsc{Bu2023Ye} simulation made with \possis\ ~for the best-fit parameters of the tight run and for an observer viewing angle of $\theta_{\rm obs}=28^\circ$. At early times ($\lesssim2$\,d), a significant fraction ($\gtrsim50$\%) of the escaping radiation originates in the squeezed-polar lanthanide-poor component. 

\bsp	
\label{lastpage}
\end{document}